# FUTURE PROSPECTS FOR STUDYING CP VIOLATION IN B-MESON DECAYS

TATSUYA NAKADA

Department of Nuclear and Particle Physics, Paul Scherrer Institute,
CH-5232 Villigen-PSI, Switzerland

## ABSTRACT

Experimental prospects for observing CP violation in B-meson decays are reviewed. Comparisons are made for various options: experiments at $e^+e^-$ B-Meson Factories, HERA and the TEVATRON will produce results in near future. They will have a good chance to discover CP violation in B-meson decays. On a longer time scale, experiments at the LHC will aim at accurate measurements to make a precision test of the standard model in CP violation.

## 1. Introduction

Together with baryon number violation, CP violation is one of the necessary conditions to generate asymmetry in the amount of matter and anti-matter in the universe[1]. Although the standard model can generate baryon number violation and CP violation, their effects are believed to be far too small for explaining the observed asymmetry[2]. Detected CP violation in the kaon system can be accommodated in the standard model. However, it is not excluded that we are already observing a sign of new physics which was responsible for the matter anti-matter asymmetry.

Experimental observation of CP violation is still limited to the neutral kaon system[3]. However, performing a precision test of the standard model in CP violation does not look possible in the kaon system. This is due to the large theoretical uncertainties in calculating the standard model predictions[4]. Decay channels which can be used to study CP violation are also limited in the kaon system. In the B-meson system, the standard model can make good predictions for CP asymmetries in several decay modes[5]. Therefore, the B-meson system appears to be the best place to make a precision study of CP violation[6]. Since the branching fractions for relevant decay modes are all small, many B-mesons are required. In this paper, we discuss how well those CP asymmetries could be measured by future experiments.

## 2. CP Violation and Standard Model

In the standard model, mixing of quark flavours are described by a three-by-three unitary matrix

$$V_{\text{CKM}} = \begin{pmatrix} V_{\text{ud}} & V_{\text{us}} & V_{\text{ub}} \\ V_{\text{cd}} & V_{\text{cs}} & V_{\text{cb}} \\ V_{\text{td}} & V_{\text{ts}} & V_{\text{tb}} \end{pmatrix} \quad (1)$$

usually referred to as the CKM matrix[7].

Moduli for seven of the nine elements, $|V_{\text{ud}}|$, $|V_{\text{us}}|$, $|V_{\text{ub}}|$, $|V_{\text{cd}}|$, $|V_{\text{cs}}|$, $|V_{\text{cb}}|$ and $|V_{\text{tb}}|$ are determined from nuclear $\beta$-decays and pion, kaon, hyperon, D-meson, B-meson and the top quark decays by assuming that the tree diagrams dominate the decay processes, i.e. the first order weak interactions. The remaining two elements related to the top quark can be accessed only through loop induced processes: these can be the first order weak interactions such as the QCD penguin diagrams which generate rare decay modes like $B_s \to \phi K_S$ and $B_d \to K\overline{K}$. Other loop induced processes are the box diagrams which explain $K^0$-$\overline{K}^0$ and $B^0$-$\overline{B}^0$ oscillations as shown in figure 1. Since the box diagrams are the second order weak interactions, some new physics with much weaker force can contribute to the processes.

From the box diagrams, the mass difference between the two $B_d(B_s)$-meson mass eigenstates, $\Delta m_{d(s)}$, and CP violation parameter in $K^0$-$\overline{K}^0$ oscillations, $\epsilon_K$ are given by[8]

$$\Delta m_d \approx \frac{G_F^2 f_{B_d}^2 B_{B_d} m_{B_d} m_W^2}{6\pi^2} \eta_{B_d} S(x_t) |V_{\text{td}}|^2 |V_{\text{tb}}|^2 \quad (2)$$

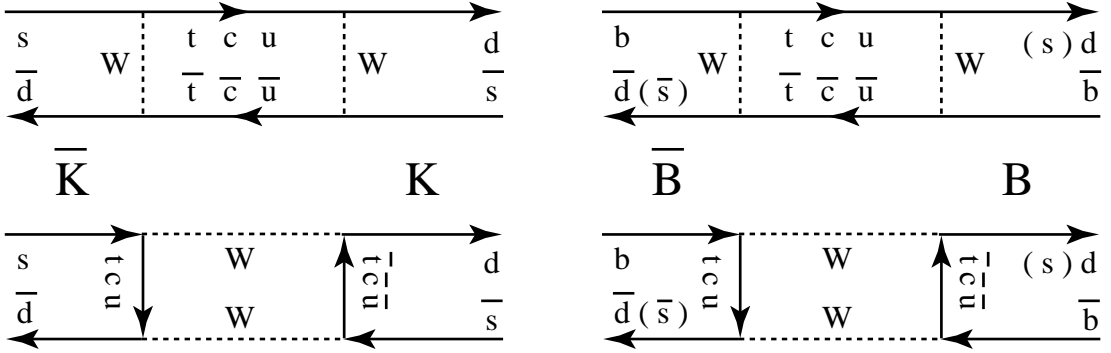

Figure 1: Box diagrams describing $K^0$-$\overline{K}^0$ and $B^0$-$\overline{B}^0$ oscillations

$$\Delta m_s \approx \frac{G_F^2 f_{B_s}^2 B_{B_s} m_{B_s} m_W^2}{6\pi^2} \eta_{B_s} S(x_t) |V_{ts}|^2 |V_{tb}|^2 \quad (3)$$

$$|\epsilon_K| \approx \frac{G_F^2 f_K^2 B_K m_K m_W^2}{12\sqrt{2}\pi^2 \Delta m_K} \operatorname{Im}\left[\eta_1 \xi_c^2 S(x_c) + 2\eta_2 \xi_c \xi_t E(x_c, x_t) + \eta_3 \xi_t^2 S(x_t)\right] \quad (4)$$

where $\xi_q = V_{qs}V_{qd}^*$, $G_F$ is the Fermi constant, $f_B$ ($f_K$), $B_B$ ($B_K$) and $m_B$ ($m_K$) are the decay constant, $B$ parameter and mass for the B(K)-meson respectively and $m_W$ is the mass of the W-boson. The QCD correction factors are denoted by $\eta_B$, $\eta_1$, $\eta_2$ and $\eta_3$ and $S$ and $E$ are known functions of the mass ratios[9], $x_i = m_i^2/m_W^2$ for top (i=t) and charm (i=c) and $\Delta m_K$ is the $K_L$-$K_S$ mass difference. Approximations are based on $x_t \approx 1$, $x_c \ll 1$ and $x_u \approx 0$.

Using the three angles and one phase, $\theta_{12}$, $\theta_{23}$, $\theta_{13}$ and $\delta$ respectively, the "standard" parameterization[3] for the CKM matrix can be given by

$$V_{KM} = R_{23} \times R_{13} \times R_{12} \quad (5)$$

where

$$R_{12} = \begin{pmatrix} c_{12} & s_{12} & 0 \\ -s_{12} & c_{12} & 0 \\ 0 & 0 & 1 \end{pmatrix}, R_{23} = \begin{pmatrix} 1 & 0 & 0 \\ 0 & c_{23} & s_{23} \\ 0 & -s_{23} & c_{23} \end{pmatrix}, R_{13} = \begin{pmatrix} c_{13} & 0 & s_{13}e^{-i\delta} \\ 0 & 1 & 0 \\ -s_{13}e^{i\delta} & 0 & c_{13} \end{pmatrix} \quad (6)$$

with $s_{ij} = \sin\theta_{ij}$ and $c_{ij} = \cos\theta_{ij}$. Wolfenstein's representation[10] can be derived by introducing

$$\lambda = \sin\theta_{12}, A = \frac{s_{23}}{s_{12}^2}, \rho = \frac{s_{13}\cos\delta}{s_{12}s_{23}}, \eta = \frac{s_{13}\sin\delta}{s_{12}s_{23}}. \quad (7)$$

The matrix is often approximated by taking into account terms up to an order of $\lambda_c^3$: i.e.

$$V_{CKM}^{(3)} = \begin{pmatrix} 1 - \lambda^2/2 & \lambda & A\lambda^3(\rho - i\eta) \\ -\lambda & 1 - \lambda^2/2 & A\lambda^2 \\ A\lambda^3(1 - \rho - i\eta) & -A\lambda^2 & 1 \end{pmatrix}. \quad (8)$$

Using the data from nuclear $\beta$-decays and pion, kaon, hyperon, D-meson and B-meson decays, we currently obtain[11]

$$\lambda = 0.2205 \pm 0.0018, \quad A = 0.794 \pm 0.054, \quad \sqrt{\rho^2 + \eta^2} = 0.363 \pm 0.101. \qquad (9)$$

The small value extracted for $\lambda$ justifies this approximation. Errors on those parameters will continue to improve in the future.

For describing CP violation, a better approximation

$$V_{\text{CKM}} \approx V_{\text{CKM}}^{(3)} + \delta V_{\text{CKM}} \qquad (10)$$

where

$$\delta V_{\text{CKM}} = \begin{pmatrix} 0 & 0 & 0 \\ -i\, A^2 \lambda^5 \eta & 0 & 0 \\ (\rho + i\, \eta)\, \lambda^5/2 & (1/2 - \rho)\, A\lambda^4 - i\, A\lambda^4 \eta & 0 \end{pmatrix} \qquad (11)$$

is needed. The parameter $|\epsilon_K|$ given by equation 4 is now derived to be

$$|\epsilon_K| \approx \frac{G_F^2 f_K^2 B_K m_K m_W^2}{6\sqrt{2}\pi^2 \Delta m_K} A^2 \lambda^6 \tilde{\eta} \left\{ x_c \left[\eta_2 \tilde{E}(x_c, x_t) - \eta_1\right] + \eta_3 S(x_t) A^2 \lambda^4 (1 - \tilde{\rho}) \right\} \qquad (12)$$

where $\tilde{\rho} = \rho(1 - \lambda^2/2)$, $\tilde{\eta} = \eta(1 - \lambda^2/2)$ and $\tilde{E}$ is a known function of $x_c$ and $x_t$. Note that the $\delta V_{\text{CKM}}$ correction to $V_{cd}$ is not negligible in the expression. Compared with the theoretical uncertainties in $B_K$, $\delta V_{\text{CKM}}$ corrections for $V_{td}$ and $V_{ts}$ can be neglected here.

Similarly for $\Delta m_d$ and $\Delta m_s$, from equations 2 and 3 we obtain

$$\Delta m_d = \frac{G_F^2 f_{B_d}^2 B_{B_d} m_{B_d} m_W^2}{6\pi^2} S(x_t) \eta_{B_d} A^2 \lambda^6 \left[(1 - \tilde{\rho})^2 + \tilde{\eta}^2\right] \qquad (13)$$

$$\Delta m_s = \frac{G_F^2 f_{B_s}^2 B_{B_s} m_{B_s} m_W^2}{6\pi^2} S(x_t) \eta_{B_s} A^2 \lambda^4 \left(1 - \lambda^2 + 2\rho\lambda^2\right). \qquad (14)$$

From the measured $\Delta m_d$[12], we obtain $(1 - \tilde{\rho})^2 + \tilde{\eta}^2 = 1.02 \pm 0.48$ where the error is completely dominated by the theoretical uncertainty in $f_{B_d}\sqrt{B_{B_d}}$ which is taken to be $200 \pm 40$ MeV[13]. Even if $f_{B_d}$ could be measured experimentally in future from the $B^\pm \to \tau^\pm \nu_\tau$ decays, the relative error on $f_{B_d}\sqrt{B_{B_d}}$ will not become less than 10%.

A better extraction of $(1 - \tilde{\rho})^2 + \tilde{\eta}^2$ can be made if $\Delta m_s$ is measured as well; i.e.

$$(1 - \tilde{\rho})^2 + \tilde{\eta}^2 \approx \frac{\Delta m_s}{\Delta m_d} \frac{m_{B_d}}{m_{B_s}} \frac{f_{B_d}^2 B_{B_d}}{f_{B_s}^2 B_{B_s}} \qquad (15)$$

where the ratio

$$\frac{f_{B_d}^2 B_{B_d}}{f_{B_s}^2 B_{B_s}} \qquad (16)$$

is much better known[13]. We could hope that $(1 - \tilde{\rho})^2 + \tilde{\eta}^2$ will be extracted with a relative error of $\sim 15\%$ once $\Delta m_s$ will be measured. Figure 2 summarizes the situation on $\sqrt{\rho^2 + \eta^2}$ and $(1-\rho)^2 + \eta^2$ measurements with the current expectation values and one standard deviation bands. It also indicates the expected improvement once $\Delta m_\mathrm{s}$ will be measured with the 15% error.

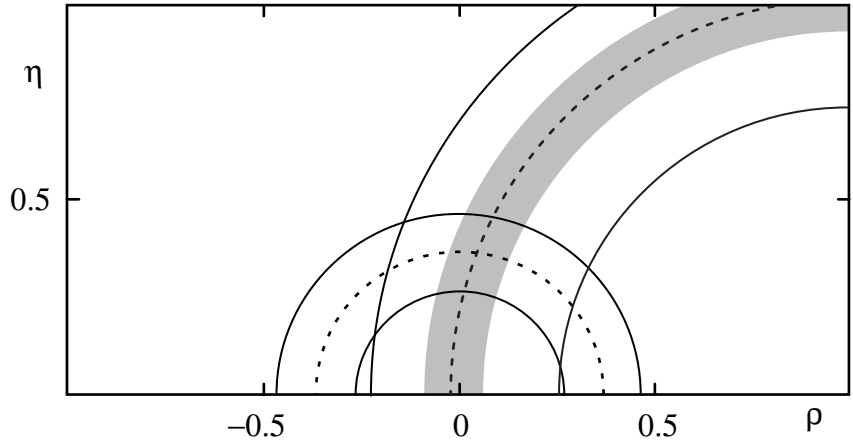

Figure 2: Expectation values and one standard deviation bands for $(\rho^2 + \eta^2)^{1/2}$ and $[(1-\rho)^2 + \eta^2]^{1/2}$ measured from the B-meson decays. Shaded band corresponds to the 15% error on $[(1-\rho)^2 + \eta^2]^{1/2}$.

The two unitarity relations relevant to CP violation in B-meson decays,

$$V_\mathrm{ud} V_\mathrm{ub}^* + V_\mathrm{cd} V_\mathrm{cb}^* + V_\mathrm{td} V_\mathrm{tb}^* = 0 \qquad (17)$$

for $B_\mathrm{d}$ and

$$V_\mathrm{tb} V_\mathrm{ub}^* + V_\mathrm{ts} V_\mathrm{us}^* + V_\mathrm{td} V_\mathrm{ud}^* = 0 \qquad (18)$$

for $B_\mathrm{s}$ in this improved approximation of $V_\mathrm{CKM}$ are illustrated in figure 3. Note that the $\delta V_\mathrm{CKM}$ correction to $V_\mathrm{cd}$ is small here and is neglected.

In summary, the following phase convention is valid for the B-meson system:

- $V_\mathrm{ud}$, $V_\mathrm{us}$, $V_\mathrm{cd}$, $V_\mathrm{cs}$, $V_\mathrm{cb}$ and $V_\mathrm{tb}$ are real.

- $\arg V_\mathrm{ub} = -\gamma$ where $\gamma = \tan^{-1} \frac{\eta}{\rho}$.

- $\arg V_\mathrm{td} = -\beta$ where $\beta = \tan^{-1} \left[ \frac{\eta}{1-\rho} \left( 1 - \frac{1}{1-\rho} \frac{\lambda^2}{2} \right) \right]$

- $\arg V_\mathrm{ts} = \delta\gamma + \pi$ where $\delta\gamma = \eta\lambda^2$

The terms proportional to $\lambda^2$ are due to the correction term $\delta V_\mathrm{CKM}$ and of the order $10^{-2}$. For an experiment at LHC, these terms should **not** be neglected.

Currently, $|\epsilon_\mathrm{K}|$ is also included when $\eta$ and $\rho$ are extracted from the data. One of such analyses gives 0.05 and 0.36[14] for the most likely values of $\rho$ and $\eta$, respectively. In future, the data from the B-meson studies alone will be sufficient for extracting $\rho$ and $\eta$: $\sqrt{\rho^2 + \eta^2}$ from $|V_\mathrm{ub}|$ and $(1-\rho)^2 + \eta^2$ from the combination of $\Delta m_\mathrm{d}$ and $\Delta m_\mathrm{s}$.

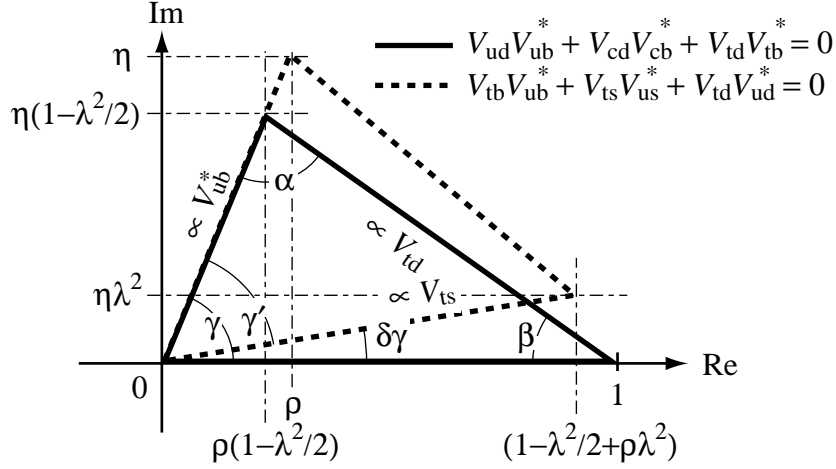

Figure 3: Unitarity triangles in the complex plane.

Three angles of the unitarity triangles are directly accessible by studying CP violation in B-meson decays. Examples are[6,15,16]

- CP asymmetry between $B^0 \to \pi^+\pi^-$ and $\overline{B}^0 \to \pi^+\pi^-$: $\implies \sin 2\alpha$ (neglecting the penguin diagrams).

- CP asymmetry between $B^0 \to J/\psi\, K_S$ and $\overline{B}^0 \to J/\psi\, K_S$: $\implies \sin 2\beta$ (penguin diagrams are negligible).

- CP asymmetry between $B_s^0 \to J/\psi\, \phi$ and $\overline{B}_s^0 \to J/\psi\, \phi$: $\implies \sin 2\delta\gamma$ (assuming that CP eigenvalues for the final states are always identical.)

- CP asymmetries between $B_s^0 \to D_s^\pm K^\mp$ and $\overline{B}_s^0 \to D_s^\mp K^\pm$: $\implies \sin\gamma'$ ($\gamma' = \gamma - \delta\gamma$).

- Measurements of six branching fractions, $B^0 \to D^0 K^{*0}, \overline{D}^0 K^{*0}, D_{1,2} K^{*0}$ and their charge conjugated decay processes where $D_{1,2}$ are CP eigenstates of the neutral D-mesons: $\implies \sin 2\gamma$ ($B^\pm$ can be used as well).

Note that the range of $\alpha$, $\beta$ and $\gamma$ is, in principle, 0 to $\pi$. Therefore, measurements sensitive to twice the angles have a twofold ambiguity. However, $\beta$ is restricted to be less than $\pi/2$ with the currently available data. Clearly one must measure all those angles using B-meson decays.

If there exists a new interaction which can generate flavour changing neutral currents, they will contribute to $K^0$-$\overline{K}^0$, $B^0$-$\overline{B}^0$ and $B_s^0$-$\overline{B}_s^0$ oscillations competing against the second order weak interaction box diagrams. Therefore, $\Delta m_d$ and $\Delta m_s$ no longer provide $(1-\rho)^2 + \eta^2$ and $|\epsilon_K|$ cannot be used for extracting $\rho$ and $\eta$. The penguin diagrams will be less affected from such a new interaction which should compete against the first order weak interactions. Therefore, $(1-\rho)^2 + \eta^2$ might be better determined from rare decays[17]. If the new interaction has a complex coupling, CP asymmetries in $B_d \to \pi^+\pi^-$, $B_d \to J/\psi\, K_S$, $B_s \to J/\psi\, \phi$ and $B_s \to D_s K$ are also affected. Thus, they do not give the angles of the unitarity triangles.

## 3. Experimental Prospects

### 3.1. General Considerations

After the discovery of the b-quark with a hadron machine at FNAL[18], properties of B-mesons were studied for long time, more or less exclusively, with $e^+e^-$ colliders running at the $\Upsilon(4S)$ resonance, CESR at Cornell and DORIS at DESY. The main advantage of experiments at the $\Upsilon(4S)$ resonance is that events are clean in the following two ways:

- One in every five hadronic events is a b-quark event.

- Only a pair of B- and $\overline{B}$-mesons is exclusively produced in one event, $\Upsilon(4S) \to B\overline{B}$, and no additional particle is present.

Other advantages are

- B-mesons are produced with a known energy of 5.29 GeV, which can be used to reduce the background in reconstructed B-mesons.

- B-mesons are produced almost at rest in the $\Upsilon(4S)$ frame; $p_B = 341$ MeV/$c$. This can be used to reconstruct the neutrino momentum in semileptonic B-meson decays.

While DORIS stopped its operation, CESR continues to run with ongoing upgrade plans for both the machine and experiment[19].

Since LEP became fully operational, studies of the b-quark have been one of the most active fields of research by the LEP experiments[20]. Advantages at LEP are:

- The b-quark cross section at the $Z^0$, $\sim 6$ nb, is higher than that at the $\Upsilon(4S)$, $\sim 1$ nb.

- In addition to $B_u$ and $B_d$, other b-hadrons such as $B_s$ and $\Lambda_b$ are produced.

- B-mesons fly a distance of a few mm in average before they decay. Therefore, decay vertices of B-mesons can be well reconstructed.

Total of $\sim 3 \times 10^6$ b-quark events have been collected by the LEP experiments. Unique achievements by the LEP experiments are the measurements of individual lifetimes for various b-hadrons, time dependent analysis of $B^0$-$\overline{B}^0$ oscillations and limits on $B_s^0$-$\overline{B}_s^0$ oscillations. Since the LEP operation has been shifted to the LEP-II programme, new data on B-mesons will no longer be available.

Although the b-quark was discovered with a hadron machine by observing narrow $\Upsilon$ resonances decaying into two muons, studies of the B-meson at hadron machines, in particular with the fixed target mode, made little progress for a long time. This was mainly due to the small $\sigma_{b\bar{b}}/\sigma_{\text{total}}$ which makes it difficult to trigger b-quark events at energies accessible by the fixed target experiments. The experimental situation is better at higher energies obtained by $p\bar{p}$ colliders. In particular, the CDF experiment equipped with a silicon micro-vertex detector is already contributing a lot to the individual lifetime measurements[21].

Before the turn of this century, two $e^+e^-$ colliders at the $\Upsilon(4S)$ now under construction, KEKB[22] at KEK and PEP-II[23] at SLAC, will become fully operational. They are often called "B-meson factories", with very high luminosities and unequal beam energies. In the same time scale, the hadron frontier will be pushed by the HERA-B[24] experiment at DESY and upgraded CDF and D0 at FNAL. All those experiments will be sensitive to physics which requires $10^8$ B-mesons. If CP violation is predominantly due to the standard model, CP violation in B-meson decays will be found in $B_d \to J/\psi K_S$ by those experiments. In the mean time, CESR will continue to run.

Around 2005, LHC will become operational. Since LHC can produce much more than $10^{10}$ B-mesons in one year, this will become the ultimate source for B physics. The goal of experiments will be to make a precision test of the standard model in CP violation by measuring CP asymmetries in different B-meson decay channels with a high accuracy. Studies of rare and forbidden B-meson decays in the standard model will also be an important part of the programme.

*3.2. Near Future*

Table 1 summarises some parameters of the two $e^+e^-$ B-meson factories. KEKB is being constructed in the existing TRISTAN tunnel and PEP-II in the PEP tunnel.

Two beams with moderately different energies will collide in both machines which will produce the $\Upsilon(4S)$ boosted. This is essential for the CP violation study. At $\Upsilon(4S)$, $B^0$-$\overline{B}^0$ are produced in a state with an orbital angular momentum of one. Due to quantum coherency of this state, one can show that the difference in the decay time between the two $B_d$-mesons must be measured in order to obtain a visible signal of CP violation. With a stationary $\Upsilon(4S)$, only the sum of the decay times can be measured since the production points of $B_d$-mesons given by the $e^+e^-$ collision points are not well defined compared with the average flight path of $B_d$-mesons. With a boosted $\Upsilon(4S)$, measuring the decay time difference becomes possible as illustrated in figure 4.

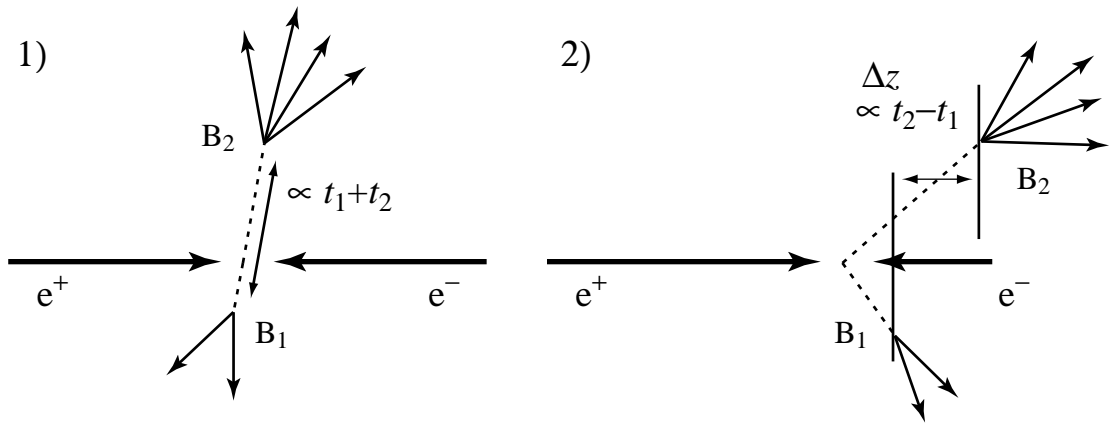

Figure 4: Illustration of $\Upsilon(4S) \to B^0\overline{B}^0$ followed by decays of the two B mesons for 1) a symmetric collider and 2) an asymmetric collider.

Table 1: List of some machine parameters for PEP-II and KEKB.

| Machine | $E_H/E_L$ [GeV] | $\mathcal{L}$ [cm$^{-2}$s$^{-1}$] | Bunch spacing [m] | Crossing angle |
|---|---|---|---|---|
| PEP-II | 9/3.1 | $3 \times 10^{33}$ | 1.26 | 0° |
| KEKB | 8/3.5 | $10^{34}$ | 0.6 | ±11 mrad |

In the PEP-II design, the two beams collide head-on. Then the two beams are separated by dipole magnets in order to avoid parasitic collisions. Separation has to be done quickly in order to increase the number of bunches, which leads to the high luminosity. For quick separation of the beams after the collision, the dipoles have to be placed very close to the interaction point, which makes the detector design difficult. In the KEK design, the two beams collide with a small angle. In this scheme, the beams are automatically separated after the collision. Therefore, no dipole is needed close to the interaction point and the bunch spacing can be reduced, i.e. the number of bunches can be increased. However, crossing bunches with a finite angle may introduce an instability of the beams which limits the luminosity.

Both machines have only one interaction region with a detector in order to obtain the highest luminosities. BELLE is placed at KEKB and BaBar at PEP-II.

Table 2: Experimental conditions for B-meson factories, HERA and the TEVATRON.

|  | Factories | HERA | TEVATRON |
|---|---|---|---|
| Experiments | BELLE    BaBar (KEKB)   (PEP-II) | HERA-B | CDF, D0 |
| Reactions | $e^+ + e^-$ at $\Upsilon(4S)$ | p+Cu at $\sqrt{s}$ =40 GeV | $p + \bar{p}$ at $\sqrt{s}$ =1.8 TeV |
| $\sigma_{b\bar{b}}$ | ∼1 nb | ∼760 nb | ∼100 μb |
| $\sigma_{b\bar{b}}/\sigma_{hadronic}$ | $\sim 2 \times 10^{-1}$ | $\sim 10^{-6}$ | $\sim 2 \times 10^{-3}$ |
| b-hadrons | $B_u$, $B_d$ | $B_u$, $B_d$, $B_s$, $B_c$, all b-Baryons | $B_u$, $B_d$, $B_s$, $B_c$, all b-Baryons |
| In one event | only $B\bar{B}$ | many others | many others |
| Geometry of detector | central, slightly asymmetric | very forward fixed target | central symmetric |
| Particle ID | p/K/π/μ/e | p/K/π/μ/e | hadron/μ/e |

The HERA-B experiment uses the halo of the HERA proton beam and internal targets placed inside the beam pipe. It runs parasitically to the other HERA experiments. The internal targets are made of eight thin Cu or Al wires (50 μm diameter). Material with a large atomic number increases the b-quark production cross section. Since it is a fixed target experiment, B-mesons are boosted with an average decay length of $\sim 6$ mm.

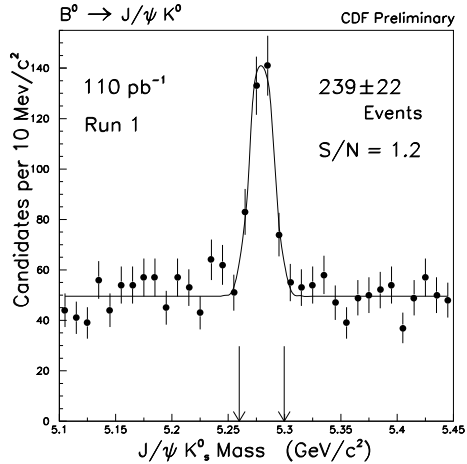

Figure 5: Invariant mass distribution for reconstructed $B_d \to J/\psi\, K_S$ decays by the CDF experiment.

Table 3: CP physics performance for BaBar, BELLE, HERA-B and CDF or D0 for one year of data taking. The parameter $x_s$ is given by $\Delta m_s/\tau_B$ where $\tau_B$ is the B-meson lifetime.

| Experiment | BaBar | BELLE | HERA-B | CDF (D0) |
|---|---|---|---|---|
| **Statistical** error on $\sin 2\alpha$ | | | | |
| $B_d \to \pi^+\pi^-$ | 0.20 | 0.15 | 0.14 | 0.10 |
| $B_d \to \rho^\pm \pi^\mp$ | 0.11 | 0.15 | - | - |
| **Statistical** error on $\sin 2\alpha$ | | | | |
| $B_d \to J/\psi K_S$ | 0.10 | 0.08 | 0.13 | 0.08 |
| $B_d \to J/\psi K_L$ | 0.16 | 0.14 | - | - |
| $B_d \to D^+D^-(D^{*+}D^{*-})$ | 0.21(0.15) | - | - | - |
| **Statistical** error on $\gamma$ | | | | |
| $B \to DK^*(K)$ | - | $12°$ | - | - |
| $B_s^0$-$\overline{B}_s^0$ oscillations: $x_s$ reach | not compatible with | | | |
| $B_s \to D_s \pi$ | CP measurements | | up to $\sim 17$ | up to $\sim 20$ |

The CDF experiment has been already demonstrated its potential by reconstructing a handful of $B_d \to J/\psi K_S$ decays[25] shown in figure 5. Both CDF and D0 are upgrading their detector which will enhance their capability for studying CP violation in B-meson decays in the next run. In particular, D0 will introduce a magnet.

Some of the experimental conditions for those three different approaches are listed in Table 2. For all the experiments, the micro-vertex detector is a crucial part of the spectrometer. It is essential for reducing the background and measuring the B-meson decay time. The second point is particularly important for BaBar and BELLE to observe CP violation.

In addition to the lepton identification, to have a capability to separate kaons from pions introduced by BaBar, BELLE and HERA-B is another important point. This can be seen for the reconstruction of the $B_d \to \pi^+\pi^-$ decay. Dangerous backgrounds in this channel are $B \to K^\pm\pi^\mp$ (also $B_s \to K^\pm\pi^\mp$ and $\to K^+K^-$ for hadron machines). Since they are real two body B-meson decays, the micro-vertex detector will not help to remove them. The mass resolutions of spectrometers are not sufficient to separate those channels using invariant masses with different mass hypotheses. Furthermore, the efficiency for the flavour tag can be increased by using the kaon tag.

BaBar and BELLE are equipped with electromagnetic calorimeters made of CsI crystals. Their excellent energy resolutions and clean environment at $\Upsilon(4S)$ decays will allow the two experiments to enhance the decay channels to be investigated by including final states with multi $\pi^0$'s. Table 3 summarises the expected performance of those experiments in CP violation studies[26].

### 3.3. Long Term Future

The $b\bar{b}$ cross section at LHC is expected to be $\sim 500~\mu$b. Even with a modest luminosity of $10^{32}$ to $10^{33}$ cm$^{-2}$s$^{-1}$, more than $10^{11}$ B-mesons will be produced in $10^7$ seconds. The cross section ratio, $\sigma_{b\bar{b}}/\sigma_{\text{inelastic}}$ is predicted to be $\sim 5 \times 10^{-3}$, which is similar to $\sigma_{c\bar{c}}/\sigma_{\text{inelastic}}$ in the current fixed target charm experiments. Therefore, LHC will be a promising machine for studying CP violation in B-meson decays with high statistics.

Three experiments are foreseen for proton-proton collisions at LHC. ATLAS and CMS are two general purpose experiments designed to look for Higgs and supersymmetric particles. LHC-B is a dedicated experiment for the CP violation study. Experimental conditions for the three experiments are summarised[27] in Table 4. All the detectors are naturally equipped with high performance micro-vertex detectors.

Since ATLAS and CMS are designed to look for particles produced in a very hard collision, detectors cover the central region. For the initial phase of the LHC operation where the machine luminosity is $\sim 10^{33}$ cm$^{-2}$s$^{-1}$, they intend to do physics with B-mesons. The b-quark events are triggered by the high transverse momentum ($p_t$) lepton trigger by reducing the threshold value.

Table 4: Comparison of ATLAS, CMS and LHC-B.

| Experiment | general purpose detectors ATLAS    CMS | dedicated experiment LHC-B |
|---|---|---|
| $\mathcal{L}$ for B physics | $10^{33}$ cm$^{-2}$s$^{-1}$ | $1.5 \times 10^{32}$ cm$^{-2}$s$^{-1}$ |
| Acceptance | central region $\|\eta\| < 2.5$     $\|\eta\| < 2.4$ | forward region $1.6 < \eta < 5.3$ |
| Level-1 trigger | high-$p_t$ $\mu$     high-$p_t$ $\mu$+e | medium-$p_t$ $\mu$, e and hadrons |
| Particle ID | hadron/$\mu$/e | p/K/$\pi$/$\mu$/e |

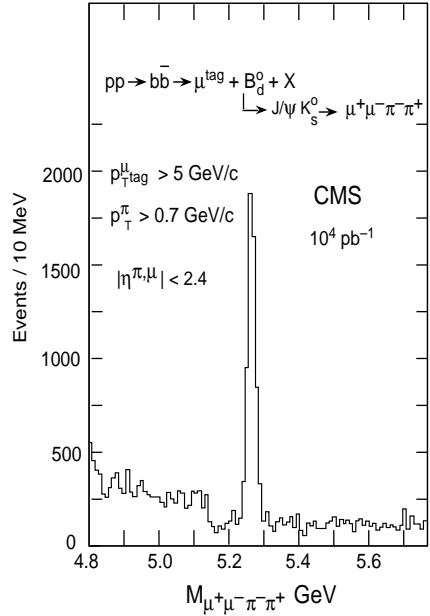

Figure 6: Simulated invariant mass distribution for $B_d \to J/\psi\, K_S$ decays for the CMS detector.

LHC-B chose the forward geometry due to the following reasons:

- The b-quark production is peaked in the forward direction and in the froward region both b and $\overline{b}$ go to the same direction. Therefore, a single arm spectrometer with a modest angular coverage of up to $\sim 400$ mrad can detect 10 to 20% of $b\overline{b}$ events where decay products of the both b-hadrons are in the detector acceptance. This reduces the cost of the detector.

- B-hadrons produced in the forward direction are faster than those in the central region. Their average momentum is about 80 GeV/$c$, corresponding to a mean decay length of $\sim 7$ mm. Therefore, a good decay time resolution can be obtained for reconstructed B-mesons.

- In the forward region, momenta are mainly carried by the longitudinal components. Therefore, the threshold value for the $p_t$ trigger can be set low for electrons, muons and hadrons; around 1.5 GeV/$c$. This makes the $p_t$ trigger more efficient than in the central region.

- The detector can be built in an open geometry which allows easy installation and maintenance.

Table 5 summarises[28] the performance of three detectors in CP violation studies. It includes expected **statistical** accuracies for measured CP violation parameters.

Table 5: Comparison of CP physics performance for ATLAS, CMS and LHC-B with $10^7$ s data taking. The branching fraction for $B_d \to \pi^+\pi^-$ is scaled to $1.2 \times 10^{-5}$.

| Experiment | ATLAS | CMS | LHC-B |
|---|---|---|---|
| $\sin 2\alpha$ using $B_d \to \pi^+\pi^-$ | | | |
| B mass resolution | 50 MeV/$c^2$ | 27 MeV/$c^2$ | 14 MeV/$c^2$ |
| Background/Signal | $\sim 1$ | $\sim 1$ | can be made to $\sim 0$ |
| **Statistical** error on $\sin 2\alpha$ | 0.05 | 0.08 | 0.05 |
| $\sin 2\beta$ using $B_d \to J/\psi K_S$ | | | |
| B mass resolution | 16 MeV/$c^2$ | 12 MeV/$c^2$ | 7 MeV/$c^2$ |
| $J/\psi$ from | $\mu^+\mu^-$ and $e^+e^-$ | $\mu^+\mu^-$ | $\mu^+\mu^-$ and $e^+e^-$ |
| **Statistical** error on $\sin 2\beta$ | 0.02 | 0.05 | 0.02 |
| $\gamma$ using $B_d \to DK^*$ and $B_s \to D_s K$ | | | |
| **Statistical** error on $\gamma$ | - | - | $6°$ to $12°$ |
| $\sin 2\delta\gamma$ using $B_s \to J/\psi\phi$ | | | |
| **Statistical** error on $\sin 2\delta\gamma$ | 0.06 ($x_s = 25$) | - | 0.02 ($x_s = 30$) |
| $B_s^0$-$\overline{B}_s^0$ oscillations using $B_s \to D_s\pi(3\pi)$ | | | |
| $B_s$ decay time resolution | 0.07 ps | - | 0.04 ps |
| $x_s$ reach | up to $\sim 37$ | up to $\sim 30$ | up to $\sim 55$ |

As already demonstrated by CDF, general purpose collider experiments can reconstruct the $J/\psi K_S$ final state well. As an example, figure 6 shows the simulated invariant mass distribution for reconstructed $B_d \to J/\psi K_S$ decays with the CMS detector.

In order to study CP violation in many different decay channels, which is vital for a CP experiment at LHC, particle identification is essential. For example the parameter $\gamma'$ can be best measured by the $B_s \to D_s K$ decay. The most important background for this decay mode is $B_s \to D_s \pi$. The branching fraction of this channel is more than 10 times larger than that of $B_s \to D_s K$. No CP violation effect is expected in the $B_s \to D_s \pi$ decay. Therefore, separation of kaon and pion is needed for reconstructing cleanly the $B_s \to D_s K$ decay and extracting $\gamma'$.

Importance of the $K/\pi$ separation can be also demonstrated in the $B_d \to \pi^+\pi^-$ decays. As discussed previously, the major background comes from other two body decay modes of the B-meson. Figure 7 shows simulated $\pi^+\pi^-$ invariant mass distributions for reconstructed $B_d$ mesons for ATLAS. The solid lines show the "observed" distributions. The dotted line is the contribution coming only from the real $B_d \to \pi^+\pi^-$ decays. The background peaking at the $B_d$-meson mass is due to other two body B-meson decays and the combinatorial background is flat under the mass peak.

At LHC-B, one can study the background since pions, kaons and protons are identified by Ring Imaging Cherenkov detectors. Figure 8 demonstrates how different particle identification cuts reject those two body decay backgrounds. The solid line is reconstructed $B_d \to \pi^+\pi^-$ decays and shaded dotted lines are contribution from the other two body B-meson decay final states. In some decay modes like $B_d \to K^\pm \pi^\mp$, a CP violation effect is expected. The size of CP asymmetry could be as large as $\sim 0.01$. Compared with the expected statistical accuracy of the measurements shown in Table 5, this effect is not negligible. Furthermore, a discovery of CP violation in $B_d \to K^\pm \pi^\pm$ decays by itself is very important. Such studies are clearly possible only with the LHC-B detector.

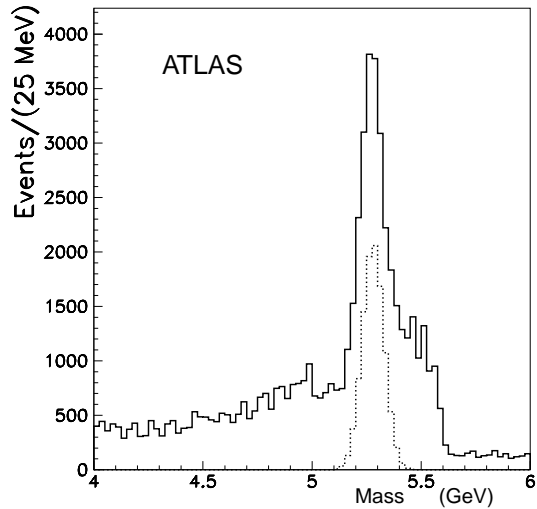

Figure 7: Simulated $\pi^+\pi^-$ invariant mass distributions for reconstructed $B_d \to \pi^+\pi^-$ decays in ATLAS. The dashed line is the contribution from the true $B_d \to \pi^+\pi^-$ decays.

LHC-B is only the detector capable of separating kaons from pions in all the necessary phase space. Thus, LHC-B can exploit the large number of B-mesons produced at LHC by well controlling the systematic effects.

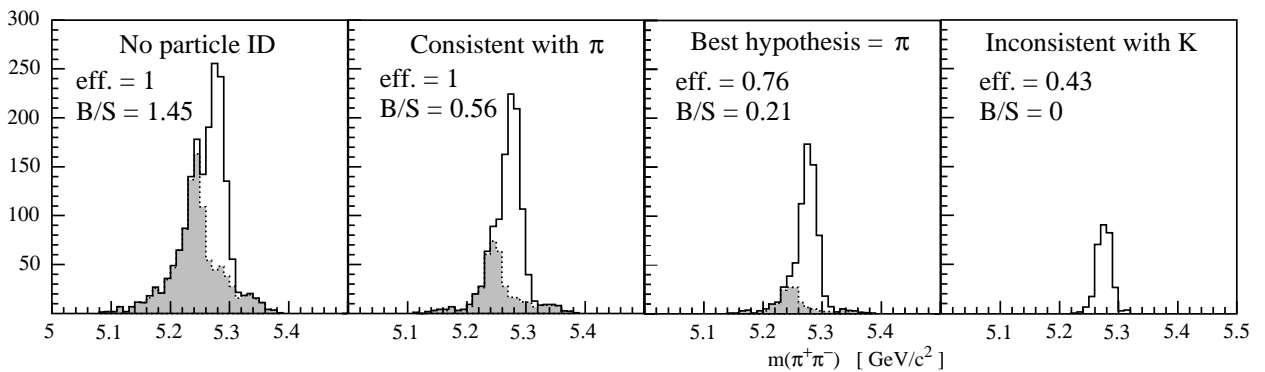

Figure 8: Simulated $\pi^+\pi^-$ invariant mass distributions for reconstructed $B_d \to \pi^+\pi^-$ decays for LHC-B with various particle identification cuts. Shaded regions are the background contributions from the other two-body decay modes of B-mesons.

It should be also noted that the $\delta V_{CKM}$ correction becomes important for measurements with sensitivities given in table 5. For demonstration, the standard model predictions of $\sin 2\alpha$, $\sin 2\beta$ and $\sin 2\delta\gamma$ are calculated for the leading order approximation $V_{\text{CKM}}$ and including the $\delta V_{\text{CKM}}$ corrections. We assume $(\rho, \eta) = (0.05, 0.36)^{14}$ in the calculations.

$$\begin{array}{ccc} & V_{\text{CKM}}{}^{(3)} & V_{\text{CKM}}{}^{(3)} + \delta V_{\text{CKM}} \\ \sin 2\alpha & 0.662 & 0.650 \\ \sin 2\beta & 0.434 & 0.418 \\ \sin 2\delta\gamma & 0 & 0.018 \end{array} \qquad (19)$$

## 4. Conclusions

There exists a solid experimental programme to study CP violation in B-meson decays which goes beyond 2005. If CP violation is predominantly produced by the standard model, CP violation in $B_d \to J/\psi K_S$ will be discovered by $\sim 2000$. However a real precision test of the standard model in CP violation will be made by the LHC experiments, in particular by LHC-B sometime after 2005. Simulation studies show that the LHC-B experiment will be able to achieve sensitivities beyond the leading order Wolfenstein approximation of the CKM matrix.

## 5. Acknowledgements

The author thanks the organisers of this workshop for their hospitality. R. Forty is acknowledged for his careful reading of this manuscript.